# Cepstral Smoothing of binary masks for convolutive blind separation of speech mixtures

*,1Ibrahim Missaoui, 2Zied Lachiri

1,2National Engineering School of Tunis, ENIT, BP. 37 Le Belvédère, 1002 Tunis, Tunisia
brahim.missaoui@enit.rnu.tn

2National Institute of Applied Science and Technology, INSAT, BP 676 centre urbain cedex, Tunis, Tunisia, zied.lachiri@enit.rnu.tn

### Abstract

*In this paper, we propose a novel separation system for extracting two speech signals from two microphone recordings. Our system combines the blind source separation technique with cepstral smoothing of binary time-frequency masks. The last is composed of two steps. First, the two binary masks are estimated from the separated output signals of BSS algorithm. In the second step, a cepstral smoothing is applied of these spectral masks in order to reduce musical noise typically produced by time-frequency masking. Experiments were carried out with both artificially mixed speech signals using simulated room model and two real recordings. The evaluation results are promising and have shown the effectiveness of our system.*

## 1. Introduction

Convolutive Blind Speech Separation (CBSS) aims at recovering a set of speech signals from several mixture signals which are modeled as linear convolutive model. In this model, each mixture signal is assumed to be a sum of filtered versions of the original signals. The classic example of this challenging is known as the cocktail party problem. It has been investigated and studied in extensive research works [1], [2].

The CBSS approaches can be classified into two categories [14]: The time-domain approach and the frequency-domain approach. The first one consist to treat the CBSS problem in time domain by directly applying the separation system to the convolutive mixtures in order to estimate the unmixing filters [3],[4], [9],[17]. It represents an extension of instantaneous domain algorithm to the convolutive mixtures models. However, the time-domain approach has drawback that requires a considerable computational cost in estimating the filter coefficients associated to the convolution operation [4],[17]. To avoid this computational complexity, the frequency-domain approach consists of transforming the CBSS problem into multiple instantaneous problems using the short time Fourier transform [5], [6]. Hence, the CBSS problem can be then treated using the instantaneous BSS approaches at each of the frequency bins.

The developed CBSS approaches can exploit the higher order statistics [8], [22] or second order statistics [18], [19]. However, there is still no reliable CBSS algorithm that can be used for the different convolutive speech mixture, especially for reverberant and noisy mixtures. Hence, in these cases, the separation performance of CBSS algorithms is still limited and need further improvement.

Recently, the notion of ideal binary mask has been proposed in computational auditory scene analysis (CASA) [15]. This novel technique based on time-frequency masking, has shown to be very well suited for speech separation. In fact, it has shown promising properties as well in suppression interference as in improving intelligibility of target signal. The binary mask can be obtained by comparing the time-frequency units of the background interference and the target speech signal, where it is labeled with a value 1 if the target energy is stronger than the interference energy and with a value 0 otherwise [13], [15]. However, the estimate of an accurate ideal binary mask from the mixtures signals is a difficult task due to the absent of clean target speech and interfering sound. The CASA approach assumes that both the target speech and interfering sound are known a priori. Nevertheless, in practice, we have only the mixture signals, what makes the estimating of an accurate ideal binary mask a difficult task.

In this work, we propose to estimate the binary masks from the separated signals obtained by using CBSS algorithm [5]. The proposed hybrid method based on the combination of BSS algorithm and the time-frequency masking can effectively address such problem associated with the use of individual methods. However, the estimate of the binary time-frequency mask is usually accompanied by some errors. These errors can lead to isolated units time-frequency and therefore give rise to the fluctuating artifacts, known as musical noise [23]. In order to overcome this problem, a cepstral smoothing procedure is applied to the estimated spectral mask. The smoothing procedure consists to perform, across different frequencies, different levels of smoothing which based on pitch information estimated from the obtained separated signals [20][21]. Indeed, it allows eliminating the unwanted isolated random peaks which lead the fluctuation artifact while preserving the broadband structures and regular pitch harmonics of speech signals [20][21][16]. The smoothed masks are converted back to the time-frequency domain and then applied to the separated signals resulting of separation operation by CBSS algorithm in order to reduce the musical artifacts.

The organization of this paper is as follows. In section 2, we state the convolutive blind speech separation problem in the frequency domain. In section 3, we describe our proposed method for the separation of two speech mixtures from two microphones. The performance of our convolutive BSS method using experimental results is shown in section 4. Finally, Section 5 concludes our work.

## 2. Basic principles of convolutive blind speech separation

We consider the convolutive case, where N source speech signals $s_i$ are mixed and observed at M sensors:

$$x_j(m) = \sum_{i=1}^{N}\sum_{p=1}^{P} h_{ji}(p) s_i(m-p+1) \qquad pour \; j=1,..,M \qquad (1)$$

With $h_{ji}$ is the impulse response of the mixing system from source *i* and sensor *j*.
This model can be written using matrix notations as:

$$X(m) = H(m) * S(m) \qquad (2)$$

Where $H(m)$ is the mixing filter matrix which is composed of impulse responses of the acoustical environment [25], $X(m) = [x_1(m),...,x_M(m)]^T$ is the vector of observation signals, $S(m) = [s_1(m),...,s_N(m)]^T$ is the vector signals and "*" is the convolution operation.

Taking a short-time Fourier transform (STFT) of the equation (1), the CBSS problem is converted to multiple instantaneous problems in the frequency domain and can be written as [5],[6]:

$$X(k,m) = H(k) S(k,m) \qquad (3)$$

Where $X(k,m)$ and $S(k,m)$ is the vector of the time-frequency representation of mixture signals and sources signals respectively.
The objective of blind speech separation is to find the unmixing filters $W(k)$ which will be used to calculate the estimated source signals according to the equation (4). In this work, we treated the case where we have two observed signals and two source signals i.e N = M = 2 and we have used the algorithm developed by Parra in [5] for estimate $W(k)$ and extracted then the sources signals:

$$\widehat{S}(k,m) = W(k)X(k,m) \qquad (4)$$

Where $\widehat{S}(k,m) = [\widehat{s}_1(k,m), \widehat{s}_N(k,m)]^T$ is the time-frequency representation of the estimated source signals.

Applying the inverse short time Fourier transform (ISTFT) to these representations, we can be obtain the time domain version of estimated signals $\widehat{s}_1$ and $\widehat{s}_2$.

## 3. Overview of the proposed CBSS system

The proposed CBSS approach, as illustrated in Figure 1, contains two modules shown in dotted boxes. In the first module, the separation of speech signals from theirs convolutive mixtures $x_1(m)$ and $x_2(m)$ is performed using the Parra's algorithm [5]. This algorithm tends to find the unmixing filters $W(k)$ in the frequency domain by performing a simultaneous diagonalization of cross-power spectrum matrices. However, a certain amount of interference is remained in the obtained separated speech signals, which will result in a degradation of sound quality. Such degradation is caused partly by the choice of length of unmixing filter and the frame length of overlap-blockshift of the STFT. Indeed, the last length is required to be broad than that of the unmixing filter. On the other hand, the unmixing filter is preferred to be sufficiently long for covering realistic reverberation.

The separation performance of Parra's algorithm is still limited and needed further improvement. In order to improve the quality of the output signals of the precedent module, we have developed a second module based on the ideal binary mask technique followed by a cepstral smoothing. The first step of this module consists on estimating two ideal binary masks $M_1(k,m)$ and $M_2(k,m)$ using the obtained separated signals $\widehat{s}_1$ and $\widehat{s}_2$. Then, a temporal smoothing of these masks is done in the cepstral domain. Finally, the obtained smoothed masks $M_1^{cepSm}(k,m)$ and $M_2^{cepSm}(k,m)$ are applied to the signals $\widehat{s}_1$ and $\widehat{s}_2$ in order to generate the final estimated signals. The description of each module is given in the following sub-sections.

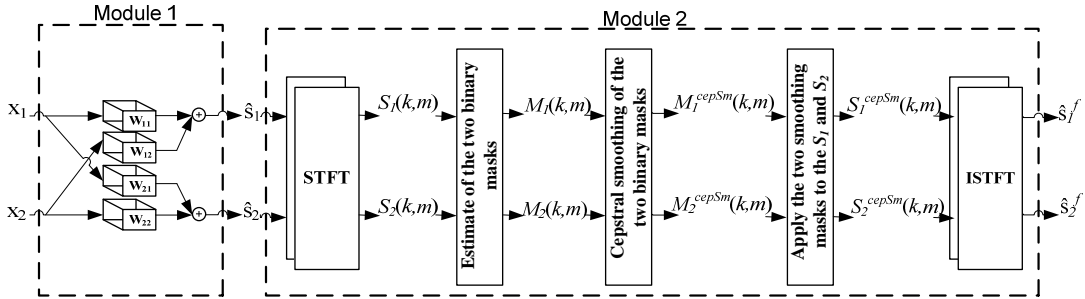

**Figure 1.** The framework of the proposed convolutive speech separation system

### 3.1. Speech separation based on non stationary signals

In the first module of our CBSS system, the Parra's algorithm proposed in [5] is applied to the signal mixtures to extract the source signals. This algorithm performs the separation task by exploiting the non stationary structure of speech which used to provide a collection of several covariance matrices to be diagonalized. The unmixing matrix $W(k)$ is found by simultaneously diagonalization of the estimated autocorrelation matrix of the speech source signals which should satisfy the following equation:

$$\Lambda_s(\omega,k) = W(\omega)\left[\bar{R}_X(\omega,k)\right]W^H(\omega) \quad (5)$$

Where $\bar{R}_X(\omega,t_k)$ is the covariance matrix of $X(\omega,t_k)$

The covariance matrices are estimated at different time using an averaged cross power spectrum. The separation criterion or the cost function, which is based on the off-diagonal elements of the matrix obtained from (6), is defined as following:

$$J = \sum_{\omega=1}^{T}\sum_{k=1}^{K}\|E(\omega,k)\|^2 \quad (6)$$

Where $E(\omega,k) = W(\omega)[\bar{R}_X(\omega,k)]W^H(\omega) - \Lambda_s(\omega,k)$

Thus, the unmixing matrix is found by minimization of the above cost function which can be solved by least squares estimation problem for different times:

$$W, \Lambda_s(\omega,k) = \underset{w(\tau)=0, \tau>Q\ll T, W_{ii}=1}{\arg\min}(J) \quad (7)$$

### 3.2. Cepstral Smoothing of binary time-frequency masks

The second module is the cepstral smoothing module. It can be divided into two steps. The first step consists of generating two ideal binary masks. The second step is to perform a cepstral smoothing.

#### 3.2.1. Ideal binary time-frequency masks

After separating the estimated speech signals $\hat{s}_1$ and $\hat{s}_2$ from their signals mixtures in the first module, the obtained signals are transformed into time-frequency domain by applying the short time Fourier (STFT) to get their spectrograms $S_1(k,m)$ and $S_2(k,m)$.

$$\begin{aligned}\hat{s}_1 &\longrightarrow S_1(k,m)\\ \hat{s}_2 &\longrightarrow S_2(k,m)\end{aligned} \quad (8)$$

Where *k* and *m* are respectively the frequency and the time frame index.
The two ideal binary masks $M_1(k,m)$ and $M_2(k,m)$ are then determined by comparing the energy of the two spectrograms at each time-frequency unit as following:

$$M_1(k,m) = \begin{cases} 1 & \text{if } |S_1(k,m)| \succ \tau|S_2(k,m)| \\ 0 & \text{otherwise} \end{cases}$$

$$M_2(k,m) = \begin{cases} 1 & \text{if } |S_2(k,m)| \succ \tau|S_1(k,m)| \\ 0 & \text{otherwise} \end{cases} \quad (9)$$

Where the threshold is chosen such $\tau = 1$.

Cepstral Smoothing of binary masks for convolutive blind separation of speech mixtures
Ibrahim Missaoui, Zied Lachiri### 3.2.2. Masks Smoothing in the cepstral domain

The previous Module is based on the time-frequency masking technique. However, this technique has the inconvenient that it usually cause fluctuating musical noise [20], [21], [23]. In order to overcome this inconvenient, we introduce a temporal smoothing step of the estimated binary masks in the cepstral domain [20]. In this step, the two obtained spectral masks $M_1(k,m)$ and $M_2(k,m)$ are transformed into cepstral representation, in which different levels of smoothing is done [20]. The idea behind employing cepstral smoothing procedure is to reduce the musical artifact produced by the time-frequency masking while preserving the broadband structure and the pitch harmonic information of the target speech signal [16], [20], [21]. Hence, this smoothing procedure should affect only unwanted random peaks. Indeed, our treatment must distinguish between broadband structures and the harmonic structure harmonics of speech on one side and unwanted random peaks on the other, which can be accomplished by applying the smoothing step in cepstral domain [16].
The cepstral representation of binary masks is given as follows:

$$M_i^{cep}(l,m) = DFT^{-1}\left\{\ln\left(M_i(k,m)\right)\big|_{k=1,..,K-1}\right\}, \quad i=1,2 \tag{10}$$

Where $l$ is the quefrency bin index, $K$ is the length of the discrete Fourier transform (DFT) respectively and "ln" is the natural logarithm [16].
Then, a first order temporal recursive smoothing is applied to the result masks $M_i^{cep}(l,m)$ as:

$$\bar{M}_i^{cep}(l,m) = \beta_l \bar{M}_i^{cep}(l,m-1) + (1-\beta_l)M_i^{cep}(l,m) \tag{11}$$

Where the value of the smoothing constants $\beta_l$ are chosen separately according to the different values of the quefrency bins $l$ as:

$$\beta_l = \begin{cases} \beta_{env} & \text{if } l \in \{0,..,l_{env}\} \\ \beta_{pitch} & \text{if } l = l_{pitch} \\ \beta_{peak} & \text{if } l \in \{(l_{env}+1),...,K\} \setminus \{l_{pitch}\} \end{cases} \tag{12}$$

Where $0 \leq \beta_{env} < \beta_{pitch} < \beta_{peak} \leq 1$, "\" is the symbol used to indicate the exclusion of $l_{pitch}$ from the quefrency rang $[l_{env}+1; K]$.
The values of the mask $M_i^{cep}(l,m)$ in the region of the low values of the quefrency bins ($0 \leq l \leq l_{env}$) represents the spectral envelope of the mask $M_i(l,m)$ [16], [20]. Since, the $\beta_{env}$ must be chosen to be a small value in order to avoid the distortion in the spectral envelope and to maintain the speech onsets. Likewise, for the quefrency bin $l = l_{pitch}$, a low smoothing $\beta_{pitch}$ is applied to this bin in order to protect the regular structure of the pitch harmonics [16], [20].
A strong smoothing (a large value for $\beta_{peak}$) is applied to the rest of quefrency bins which represent the fine structure of $M_i(l,m)$. In this range, the unwanted random peaks that generally lead to the harmonic distortion are appeared with a high probability. Consequently, these peaks are strongly affected by applying a large value of $\beta_{peak}$ in the smoothing equation (12).
The pitch frequency $l_{pitch}$ is calculated for each time-frame $m$ using the separated speech signals $\hat{s}_1$ and $\hat{s}_2$ as the following [21]:

536

$$l_{pitch} = \arg\max_l \left\{ sig^{cep}(l,m) \mid l_{low} \leq l \leq l_{high} \right\} \quad (13)$$

Where $sig^{cep}(l,m)$ is the cepstral domain representation of the estimated signal obtained in the first module. The values of $l_{low}$ and $l_{high}$ are chosen so that possible pitch frequencies 50 to 500 Hz of human speech may be accommodated.

The mask $\bar{M}_i^{cep}(l,m)$ is then determined by remaining symmetric half of cepstrum:

$$\bar{M}_i^{cep}(l,m) = \bar{M}_i^{cep}(K-l,m) \quad \text{for} \quad l \geq K/2 \quad (14)$$

The final smoothed spectral mask $M_i^f(l,m)$ is obtained as follows:

$$M_i^{cepSm}(k,m) = \exp\left( DFT\left\{ \bar{M}_i^{cep}(l,m)\mid_{l=0,\ldots,K-1} \right\} \right) \quad (15)$$

These two obtained smoothed masks are then applied to the time-frequency representation $S_1(k,m)$ and $S_2(k,m)$ of the two outputs estimated signals $\widehat{s}_1$ and $\widehat{s}_2$.

$$S_i^{cepSm}(k,m) = M_i^{cepSm} S_i(k,m) \quad (16)$$

Finally, the final estimated speech signals $\widehat{s}_1^f$ and $\widehat{s}_2^f$ are subsequently recovered in the time domain using inverse short time Fourier transform (ISTFT).

## 4. Experimental results and evaluations

To evaluate the performance of the proposed CBSS approach, we used both artificially mixed speech signals and real room recordings. The experiments were done using two convolutive mixtures consisting of two speech signals. The parameter settings for the proposed CBSS algorithm are illustrated in the table 1.

**Table 1.** Parameter values for our experiments

| DFT length= 2048 | $B_{env}=0$ | $l_{env}=8$ |
|---|---|---|
| overlap factor=0.75 | $B_{pitch}=0.9$ | $l_{low}=16$ |
|  | $\beta_{peak}=0.4$ | $l_{high}=120$ |

The performance of our system is evaluated using different performance metrics including objective and subjective evaluations. We have chosen the signal to interference ratio (SIR) as the objective evaluation. The SIR measures are generated using the BSS evaluation toolbox [7]. It is defined as follows:

$$SIR = 20\log \frac{\|s_{target}\|^2}{\|s_{interf}\|^2} \quad (17)$$

Where $s_{target}$ and $s_{interf}$ are, respectively, an allowed deformation of the target source $s_i$ and an allowed deformation of the sources which takes account of the interference of the unwanted sources.

In addition, the separated speech signals are evaluated with the Perceptual Evaluation of Speech Quality(PESQ). The PESQ, which is defined in the ITU-T P.862 standard, represents the equivalent of the subjective measure Mean Opinion Score (MOS) [24].

The experiment results of the proposed CBSS approach has been compared with that of the Parra's algorithm [5].

**4.1. Artificial convolutive mixtures**

In the first experiment, we tested our CBSS approach with different convolutive speech mixtures. These mixtures are generated by mixing two speech signals using a simulated room model with different reverberation time [25]. The reverberation time (RT) was set to 30, 50, 100, 150, 200 ms, respectively. The used speech signals, which have approximately the same loudness level, are sampled at 10 KHz and are five seconds long [12].

Table 2 presents the evaluation results of the series of experiments for artificial convolutive mixtures. It reports the SIR and PESQ measures obtained after separation by the proposed method (PM) and Parra's algorithm. As can be seen, our CBSS approach yields good results for different values of RT. The best performance was obtained for the smallest value of RT and the performance of the proposed system was degraded gradually by increasing the RT from 30 to 200. However, these results are expected due to the augmentation of sound reflections for higher reverberation.

It can be observed also that the proposed CBSS system was yielded interesting performance improvement in comparison with Parra's algorithm [5]. For example, the obtained SIR average, where the simulated room RT=30, is 20.87 db for Parra's algorithm and 31.04 db for the proposed method.

The efficiency of our CBSS system has been proved with the Perceptual Evaluation of Speech Quality (PESQ) measures. The latter was employed to measure the quality of the obtained separated speech signals. It is considered as one of the reliable methods of subjective test. The PESQ measurement returns a score from 0.5 to 4.5. As shown in table 2, we observed an improvement in speech quality of the separated signals with our CBSS system over that the Parra's algorithm. In fact, the PESQ measures confirm the obtained results in SIR and prove that using cepstral smoothing yields remarkable results as compared with the Parra's algorithm. For instance, in the experiment where the simulated room RT=30, we obtained PESQ equal to 2.93 for the proposed method and 2.83 for the Parra's algorithm.

**Table 2.** The obtained SIR and PESQ for different RT

| RT (ms) | | SIR (dB) | | PESQ | |
|---|---|---|---|---|---|
| | | Parra's algorithm | Proposed Method | Parra's algorithm | Proposed Method |
| 30 | Signal 1 | 20.75 | 26.68 | 2.83 | 2.93 |
| | Signal 2 | 20.99 | 36.13 | 3.27 | 3.42 |
| | Average | 20.87 | 31.04 | 3.05 | 3.67 |
| 50 | Signal 1 | 21.08 | 26.88 | 2.57 | 2.62 |
| | Signal 2 | 17.93 | 29.15 | 3.22 | 3.34 |
| | Average | 19.50 | 28.01 | 2.89 | 2.98 |
| 100 | Signal 1 | 12.66 | 20.78 | 1.94 | 1.94 |
| | Signal 2 | 17.61 | 27.54 | 2.79 | 2.90 |
| | Average | 15.13 | 24.16 | 2.36 | 2.42 |
| 150 | Signal 1 | 13.83 | 29.10 | 1.71 | 1.68 |
| | Signal 2 | 2.33 | 8.64 | 2.50 | 2.65 |
| | Average | 8.02 | 18.87 | 2.10 | 2.16 |
| 200 | Signal 1 | 3.72 | 17.29 | 1.60 | 1.66 |
| | Signal 2 | -0.72 | 7.51 | 2.36 | 2.42 |
| | Average | 1.5 | 12.4 | 1.98 | 2.04 |

### 4.2. Real room recordings

In this second experiment, we examine the performance of CBSS system with two practical test recordings in a real room, which were provided to the delegates of the ICA'1999 Workshop [26]. The convolutive mixtures were recorded with omnidirectional microphones in a room with dimensions of about 3.4 x 3.8 x 5.2 meters (Height x Width x Depth). These two recordings consist of two speech signals, which were produced simultaneously by two male speakers. Their sampling frequency was 16 kHz and their length is 10 seconds (160000 samples).

As the speech sources are unknown, we propose to evaluate the proposed system with the signal to interference ratio (SIR) which is estimated using the approach defined in [10]. The SIR ratio for each of the two separated speech outputs is defined as follows [10], [11]:

$$SIR_1 = 10\log \frac{p_{s_{11}}}{p_{s_{12}}}$$
$$SIR_2 = 10\log \frac{p_{s_{22}}}{p_{s_{21}}}$$
(18)

Where $p_{s_{ij}} = \sum_{n \in T_j} s_{ij}^2(n)/T_j$ is the average power and the time intervals $T_1$ corresponding to the interval where the waveform of output signal 1 has a peak and the output signal 2 exhibits low (silent) level as shown in the figure 2. The obtained segment of samples in outputs signals 1 and signal 2 are denoted as $s_{11}$ and $s_{21}$. Similarly, $T_2$ is the time interval which is characterized by the presence of the peak $s_{22}$ in waveform of output signal 1 and a lower samples (silent) $s_{21}$ in output signal 2.

As depicted in table 3, we can observe that the proposed CBSS system gives good performance for real room recordings. The SIRs of the separated speech signals produced by the proposed method are 16.36 and 14.24 dB dB respectively. The results show that our CBSS system is more effective than Parra's algorithm. For instant, we have obtained SIR average equal to 15.30 dB for the proposed method and 11.75 dB for Parra's algorithm.

**Table 3.** The obtained SIR using Parra's algorithm and the proposed method

| SIR | Parra's algorithm | Proposed Method |
|---|---|---|
| Signal 1 | 12.62 | 16.36 |
| Signal 2 | 10.88 | 14.24 |
| Average | 11.75 | 15.30 |

Cepstral Smoothing of binary masks for convolutive blind separation of speech mixtures
Ibrahim Missaoui, Zied Lachiri
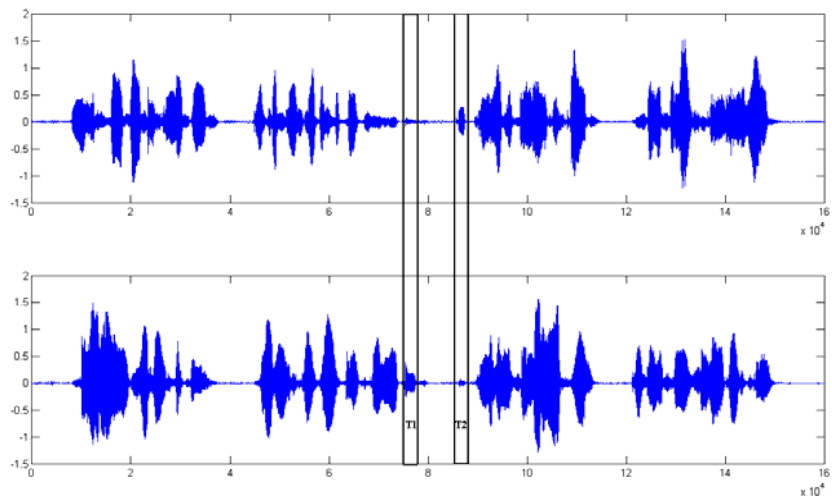

**Figure 2.** The two segments T1 and T2 of the obtained separated speech signals

## 5. Conclusion

In this work, we proposed a blind separation system of speech signals from theirs convolutive mixtures. The proposed system is based on the combination of blind source separation technique and time-frequency masking procedure, followed by a smoothing cepstral. After separating the mixture signals using CBSS algorithm, two ideal binary masks are estimated from the result separated signals. In order to reduce the musical artifact, a temporal smoothing procedure is then applied to these masks in the cepstral domain. The experimental results showed the efficiency of our system for separating two speech signals from two microphone recordings.

## 10. References

[1] S. Haykin, Z. Chen, "The cocktail party problem", Neural Computation, MIT Press Journals, vol.17, pp.1875-1902, 2005.
[2] H. Asari, B.A. Pearlmutter, A.M. Zador, "Sparse Representations for the Cocktail Party Problem", The Journal of Neuroscience, vol.26, no.28, pp.7477-7490, 2006.
[3] Ying Gao, Hongjie Li, Shifeng Ou, "Improved Momentum Term Based Adaptive Blind Source Separation Algorithm", IJACT: International Journal of Advancements in Computing Technology, vol.4, no.15, pp.12-21, 2012.
[4] S.C. Douglas, X. Sun, "Convolutive blind separation of speech mixtures using the natural gradient", Speech communication, vol.39, no.1-2, pp. 65-78, 2003.
[5] L. Parra, C. Spence, "Convolutive blind separation of non-stationary sources", IEEE Trans. on Speech and Audio Processing, vol.8, no.3, pp.320-327, 2000.
[6] S., Makino, H., Sawada, R., Mukai, S., Araki, "Blind source separation of convolutive mixtures of speech in frequency domain", IEICE Transactions on Fundamentals of Electronics, Communications and Computer Sciences E88-A, vol.7, pp.1640-1655, 2005.
[7] E. Vincent, R. Gribonval, C. Fevotte, "Performance Measurement in Blind Audio Source Separation", IEEE Transactions on Audio Speech and Language Processing, vol.14, no.4, pp.1462-1469, 2006.
[8] D. Yellin, E. Weinstein, "Multichannel signal separation: methods and analysis", IEEE Transactions on Signal Processing, vol.44, pp.106-118, 1996.
[9] Dr. Eng. Sattar, B. Sadkhan, "A Comparative Analysis of Different BSS Algorithms Based on Neural Network", IJACT: International Journal of Advancements in Computing Technology, vol.4 , no.15, pp.113-122, 2012.
540

[10] T. Mei, J. Xi, F. Yin, A. Mertins, J.F. Chicharo, "Blind source separation based on time-domain optimization of a frequency-domain independence criterion", IEEE Transactions Audio Speech Language Process, vol.14, no. 6, pp.2075-2085, 2006.
[11] T. Mei, A. Mertins, F. Yin, J. Xi, J.F. Chicharo, "Blind source separation for convolutive mixtures based on the joint diagonalization of power spectral density matrices", Signal Processing, vol.88, no.8, pp.1990-2007, 2008.
[12] M.S. Pedersen, D.L. Wang, J. Larsen, U. Kjems, "Two-microphone separation of speech mixtures", IEEE Transactions on Neural Networks., vol. 19, pp.475-492, 2008.
[13] D.L. Wang, "On ideal binary mask as the computational goal of auditory scene analysis", Speech Separation by Humans and Machines. Springer, Heidelberg, 2005.
[14] M.S., Pedersen, J. Larsen, U. Kjems, L.C. Parra, "A survey of convolutive blind source separation methods", Springer Handbook of Speech Processing, New York, 2007.
[15] D.L., Wang, D.L. Brown, "Computational Auditory Scene Analysis: Principles, Algorithms, and Applications", Wiley-IEEE Press, Hoboken, New Jersey, 2006.
[16] A.V. Oppenheim, R.W. Schafer, "Discrete Time Signal Processing", Third Edition. Prentice Hall, New Jersey, 2009.
[17] R. Aichner, H. Buchner, S. Araki, S. Makino, "On-line time-domain blind source separation of non stationary convolved signals", International Symposium on Independent Component Analysis and Blind Signal Separation, pp. 987-992, 2003.
[18] K., Rahbar, J., Reilly, "Geometric optimization methods for blind source separation of signals", International Workshop on Independent Component Analysis and Signal Separation, pp.375-380, 2000.
[19] D. Chan, P. Rayner, S. Godsill, "Multi-channel signal separation", Proceedings of the IEEE International Conference on Acoustics, Speech, and Signal Processing, pp.649-652, 1996.
[20] N. Madhu, C. Breithaupt, R. Martin, "Temporal smoothing of spectral masks in the cepstral domain for speech separation", In Proceedings of the IEEE International Conference on Acoustics, Speech and Signal Processing, pp. 45-48, 2008.
[21] T. Jan, W. Wang, D.L. Wang, "A multistage approach for blind separation of convolutive speech mixtures", IEEE International Conference on Acoustics, Speech and Signal Processing, pp.1713-1716, 2009.
[22] J. Pesquet, B. Chen, A.P. Petropulu, "Frequency domain contrast functions for separation of convolutive mixtures", Proceedings of the IEEE International Conference on Acoustics, Speech, and Signal Processing, pp.2765-2768, 2001.
[23] S. Araki, S. Makino, H. Sawada, R. Mukai, "Reducing musical noise by a fine-shift overlap-add method applied to source separation using a time-frequency mask". 2005. Proceedings of IEEE International Conference on Acoustics, Speech, and Signal Processing, pp.81-84. 2005.
[24] ITU-T P.862, "Perceptual evaluation of speech quality (PESQ), an objective method for end-to-end speech quality assessment of narrow-band telephone networks and speech codecs", International Telecommunication Union, Geneva, 2001.
[25] N.D. Gaubitch, "Allen and Berkeley image model for room impulse response", Imperial College London, 1979.
[26] ICA'99 Evaluation Data. Workshop on independent component analysis and blind signal separation, 1999. http://www2.ele.tue.nl/ica99//realworld2.html. Case 1B